\documentclass[10pt, conference, twocolumn]{IEEEtran}                                                  % paper

\IEEEoverridecommandlockouts                              % This command is only
\title{Lossless Compression with Security Constraints}

\author{%
  \authorblockN{Deniz G\"{u}nd\"{u}z\authorrefmark{1}\authorrefmark{2},
    Elza Erkip\authorrefmark{3}\authorrefmark{1},
    H. Vincent Poor\authorrefmark{1}
  }\\
  \authorblockA{%
    \authorrefmark{1}Dept.\ of Electrical Engineering,
                     Princeton University, Princeton, NJ, 08544\\
  }
    \authorblockA{%
    \authorrefmark{2}Dept.\ of Electrical Engineering,
                     Stanford University,Stanford, CA, 94305\\
  }
  \authorblockA{%
    \authorrefmark{3}Dept.\ of Electrical and Computer Engineering,
                     Polytechnic University, Brooklyn, NY, 11201 \\
  }
  Email: \{dgunduz, poor\}@princeton.edu, elza@poly.edu
  \thanks{This research was supported in part by the US National Science Foundation under Grants ANI-03-38807, CCF-04-30885, CCF-06-35177, CCF-07-28208, and CNS-06-25637.}
}

\date{December, 2006}
\usepackage{amsfonts}
\usepackage{amssymb}
\usepackage{psfrag}
\usepackage{amsmath}
\usepackage{amscd}
\usepackage{graphicx}

\graphicspath{{fig/}{num/}}

\newtheorem{thm}{Theorem}[section]
\newtheorem{cor}[thm]{Corollary}
\newtheorem{lem}[thm]{Lemma}

\newtheorem{defn}{Definition}[section]

\begin{document}
%\fontsize{12}{12} \selectfont
\maketitle \thispagestyle{empty}
\pagestyle{empty}

%%%%%%%%%%%%%%%%%%%%%%%%%%%%%%%%%%%%%%%%%%%%%%%%%%%%%%%%%%%%%%%%%%%%%%%%%%%%%%%%
\begin{abstract}
Secure distributed data compression in the presence of an eavesdropper
is explored. Two correlated sources that need to be reliably transmitted to a legitimate receiver are available at separate encoders. Noise-free, limited rate links from the encoders to the legitimate receiver, one of which can also be perfectly observed by the eavesdropper,
are considered. The eavesdropper also has its own correlated observation. Inner and outer bounds on the achievable compression-equivocation rate region are given. Several different scenarios involving the side information at the transmitters as well as multiple receivers/eavesdroppers are also considered.

\end{abstract}

\section{Introduction}

With the emergence of wireless sensor networks and distributed video applications, distributed source compression has become an important research area. A significant amount of effort has been devoted to understanding the information theoretic limits of distributed lossless and lossy compression and developing codes to achieve these limits. However, in many real-life applications involving distributed compression, such as distributed video surveillance or monitoring of some private information, secure compression and communication while meeting the end-to-end quality of service requirements becomes important. In this paper we consider the information theoretic limits of secure lossless source compression in the presence of an adversary who has access to some  of the links in the network as well as its own correlated observation of the data to be compressed. We consider information theoretic secrecy, that is, we want to limit the information leakage to a computationally unbounded eavesdropper who has the full knowledge of the compression algorithms used.

We first consider a simplified model of the general secure distributed compression problem, composed of two transmitters Alice and Charlie with correlated observations, a receiver Bob, and an eavesdropper Eve who is interested in the data of Alice. Eve eavesdrops Alice's channel to Bob, i.e., it knows Alice's message to Bob exactly. Eve also has her own correlated side information. We consider the scenario in which both Alice's and Charlie's data need to be reconstructed at Bob reliably while Eve is interested in only Alice's information source. Later, we consider various cases involving the availability of the side information at different terminals. Finally, we analyze cases with multiple receivers or multiple eavesdroppers.

%Research on secure communication over noisy channels in the presence of a eavesdropper has attracted considerable recent attention. Information theoretic security in this context is defined through the equivocation rate at the eavesdropper, which can be roughly defined as the uncertainty of the eavesdropper about the message after observing the channel output. In his pioneering work \cite{Wyner}, Wyner introduced the wire-tap channel, and showed that it is possible to transmit at a positive rate with perfect secrecy, assuming the eavesdropper's channel is physically degraded with respect to the receiver. This result is extended to more general broadcast channels in \cite{Csiszar_Korner}.
%Various extensions of the wiretap channel model to multiuser scenarios and fading channels have recently been investigated \cite{Ender}-\cite{Lifeng}.

\psfrag{C2}{$C^N$}
\psfrag{A2}{$A^N$}
\psfrag{RC2}{$R_C$}
\psfrag{RA2}{$R_A$}
\psfrag{E2}{$E^N$}
\psfrag{DEL2}{$\Delta$}
\psfrag{AC2}{$(\hat{A}^N, \hat{C}^N)$}
\psfrag{Alice2}{Alice}
\psfrag{Bob2}{Bob}
\psfrag{Eve2}{Eve}
\psfrag{Charles2}{Charlie}
%----------------------------------------------
\begin{figure}
\centering
\includegraphics[width=2.9in]{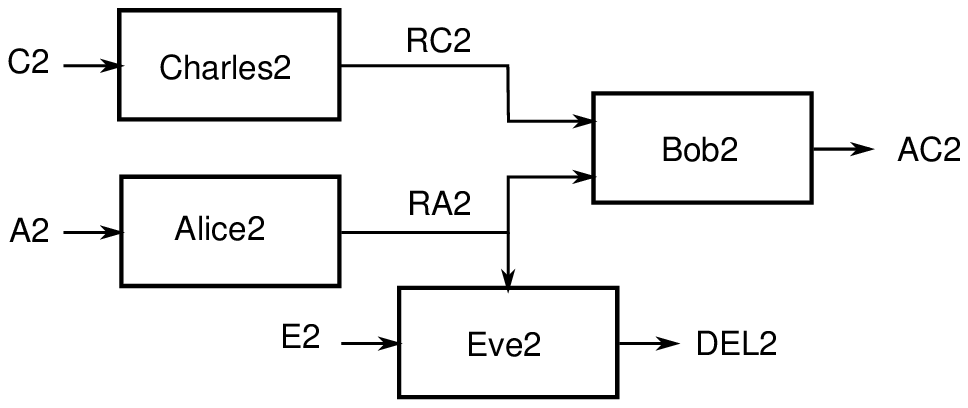}
\caption{Two terminal secure distributed compression. The eavesdropper (Eve) can only access one of the links.}
\label{f:codedSI}
\end{figure}
%----------------------------------------------

In Wyner's classical wiretap channel model \cite{Wyner}, nonzero secrecy rate can be achieved without using a secure key, if the intended receiver has a better quality communication channel than the eavesdropper. It was observed in \cite{Ahlswede_Csiszar} and \cite{Maurer} that, secrecy can also be generated through correlated observations at the legitimate users. In our model, since the channels are not noisy, the techniques of \cite{Wyner} do not apply; however, based on the ideas of \cite{Ahlswede_Csiszar, Maurer}, it is still possible to achieve secrecy by exploiting the correlated information transmitted over secure links. Unlike \cite{Ahlswede_Csiszar, Maurer} which focus on generating secret key using correlated information sources, we impose the requirement of lossless decoding of the source sequence at the legitimate receiver while keeping Alice's information secret from Eve.

In \cite{Yamamoto}, Yamamoto considers lossy compression with security constraints over a noisy broadcast channel, while the users share a secure key as well. He showed that first applying lossy source compression, then encrypting the compressed bits using the secure key and finally transmitting over the channel using a good wiretap channel code is optimal. In  \cite{Merhav}, Merhav extends this result to the case in which the legitimate receiver and the eavesdropper have correlated side information under the assumption that both the channel output and the side information at the eavesdropper are physically degraded. He shows that replacing lossy compression with Wyner-Ziv compression in the coding scheme of \cite{Yamamoto} is optimal. In \cite{Ramc}, the minimum leakage rate in secure lossless compression with arbitrary side information is explored. It is shown in \cite{Ramc} that, in the case of arbitrarily correlated receiver side information, the usual Slepian-Wolf compression is not always sufficient. Secure lossless compression of two correlated sources is considered in \cite{Kundur}, where the eavesdropper has access to only one of the compressed bit streams and has no side information. Slepian-Wolf compression suffices in this setup due to the lack of side information at the eavesdropper.

We introduce the system model in Section \ref{s:system}. In Section \ref{s:SW_coded} we give inner and outer bounds to the achievable compression-equivocation rate region that generalize the well-known Slepian-Wolf region to include secrecy constraints. In Section \ref{s:uncoded_SI} we consider various different scenarios based on the availability of the side information and also considering multiple legitimate receivers or multiple eavesdroppers.

\section{System Model}\label{s:system}

For the model in Fig. \ref{f:codedSI}, we assume that Alice and Charlie have access to length-$N$ correlated source sequences $A^N$ and $C^N$, respectively. They want to transmit these sources to Bob reliably over separate noise-free, finite capacity channels. Alice's transmission will also be perfectly received by an eavesdropper called Eve who has her own correlated side information $E^N$. We model $A^N$, $C^N$, and $E^N$ as being generated independent and identically distributed (i.i.d.) according to the joint probability distribution $p_{ACE}(a,c,e)$ over the finite alphabet $\mathcal{A} \times \mathcal{C} \times \mathcal{E}$. While Alice and Charlie want to transmit their sources reliably to Bob, they also want to maximize the equivocation at Eve, which represents the uncertainty of Eve about $A^N$ after receiving Alice's transmission and combining with her (Eve's) own side information $E^N$. We will also consider scenarios involving multiple legitimate receivers/eavesdroppers for which similar definitions apply. Throughout the paper we assume that all the transmissions are authenticated, i.e., the eavesdropper is passive.

%\begin{figure}
%\center
%  \includegraphics[scale=.5]{sec_uncodedSI.eps}
%  \caption{Illustration of the broadband relay channel }\label{bbrelaychannel}
%\end{figure}

An $(M_A,M_C,N)$ code for secure source compression in this setup is composed of an encoding function\footnote{We assume deterministic coding in the analysis for simplicity, but the proofs follow similarly for randomized coding which is modeled by assuming independent random variables at the terminals and deterministic coding functions that depend on these random variables.} at Alice, $f_A: \mathcal{A}^N \rightarrow I_{M_A} $, an encoding function at Charlie, $f_C: \mathcal{C}^N \rightarrow I_{M_C}$, and a decoding function at Bob, $g: I_{M_A} \times I_{M_C} \rightarrow \mathcal{A}^N \times \mathcal{C}^N$, where $I_{k}$ denotes the set $\{1,\ldots,k\}$ for $k \in \mathbb{Z}^+$. The equivocation rate of this code is defined as $\frac{1}{N} H(A^N|f_A(A^N), E^N)$, and the error probability as $P_e^{N} = Pr\{g(f_A(A^N), f_C(C^N)) \neq (A^N, C^N)\}$.

\begin{defn}\label{d:triplets}
We say that $(R_A, R_C, \Delta)$ is \emph{achievable} if, for any $\epsilon>0$, there exist an $(M_A,M_C,N)$ code such that $\log(M_A) \leq N(R_A + \epsilon)$, $\log(M_C) \leq N(R_C + \epsilon)$, $H(A^N|f_A(A^N), E^N) \geq N (\Delta-\epsilon)$ and $P_e^{N} < \epsilon$. Let $\mathcal{R}$ denote the set of all achievable $(R_A, R_C, \Delta)$ triplets.
\end{defn}

\section{Secure Distributed Compression}\label{s:SW_coded}

For the model in Section \ref{s:system}, when we remove the secrecy requirements, the problem reduces to the well-known Slepian-Wolf coding of correlated sources. However, the solution in the case of distributed compression with secrecy constraints is not a direct extension of the Slepian-Wolf theorem.

\begin{defn}\label{d:P_in_out}
Let $U$ and $V$ be two random variables jointly distributed with $A, C$ and $E$ and taking values over the finite alphabets $\mathcal{U}$ and $\mathcal{V}$. We define $\mathcal{P}_{in}$ as the set of $(U,V)$ that satisfy $H(C|A,V)=0$ with a joint distribution of the form $p_{ACE}p_{U|A}p_{V|C}$.  We define $\mathcal{P}_{out}$ as the set of $(U,V)$ that satisfy $H(C|A,V)=0$ and the Markov chain conditions $U-A-(C,E)$ and $V-C-(A,E)$.
\end{defn}

\begin{defn}
We define $\mathcal{R}_{in}$ as the convex hull of the set of all $(R_A, R_C, \Delta)$ for which there exists $(U,V) \in \mathcal{P}_{in}$ such that
\begin{eqnarray}
  R_C &\geq& I(C;V), \label{SWrateC}\\
  R_A &\geq& H(A|V), \label{SWrateA}\\
  \Delta &\leq& \big[I(A;V|U) - I(A;E|U)\big]^+, \label{SWrateD}\\
  \Delta &\leq & \min \{R_C - H(C|A), I(A;C) \} {~\rm and} \label{SWrateZ}\\
  \Delta &\geq & [H(A|E) - R_A]^+, \label{SWrateT}
\end{eqnarray}
where $[x]^+ = \max\{x,0\}$, and $\mathcal{R}_{out}$ as the convex hull of the set of all $(R_A, R_C, \Delta)$ for which there exists $(U,V) \in \mathcal{P}_{out}$ such that (\ref{SWrateC})-(\ref{SWrateT}) hold.
\end{defn}

Our main result is the following inner and outer bounds on the set of all achievable triplets of Definition \ref{d:triplets}.

\begin{thm}\label{t:SW_coded}
$\mathcal{R}_{in} \subseteq \mathcal{R} \subseteq \mathcal{R}_{out}$.
\end{thm}
\begin{proof}
The sketch of the proof is given in Appendix \ref{a:SW_coded}.
\end{proof}

When the eavesdropper has no side information, i.e., $E$ is constant, then the inner and outer bounds become tight. The following corollary can be obtained similar to Theorem \ref{t:SW_coded}.

\begin{cor}
When there is no side information at the eavesdropper, the compression-equivocation rate region is characterized by
\begin{align}
  R_C &\geq H(C|A), \nonumber \\
  R_A &\geq H(A|C), \nonumber \\
  R_A + R_C &\geq H(A,C),  {~\rm and} \nonumber  \\
[H(A) - R_A]^+ \leq & \Delta  \leq  \min \{I(A;C), R_C - H(C|A) \}. \nonumber
%  R_A + \Delta &\geq & ,\nonumber
\end{align}
\end{cor}
We skip the proof, which can also be obtained as a special case of the result in \cite{Kundur}. The achievability is simply from the usual Slepian-Wolf compression, that is, we let $U$ be constant.

%\begin{align}
%  R_C &\geq I(C;V) = H(C) - H(C|V) \nonumber \\
%    & =  H(C) - H(C|A, V) - I(C;A|V) \nonumber \\
%    & =  H(C) - I(C, V;A) + I(A;V) \nonumber \\
%    & \geq  H(C) - I(C;A) = H(C|A); \nonumber   \\
%  R_A &\geq H(A|V) \geq H(A|C);  \nonumber \\
%  R_A + R_C &\geq  I(C; V) + H(A|V) \nonumber \\
%        &=  H(C) - H(C|V) + H(A|V) \nonumber \\
%        &=  H(C) - H(C|A,V) - I(C;A|V) + H(A|V) \nonumber \\
%        &\geq  H(C) + H(A|V, C) = H(A,C); \nonumber \\
% \Delta &\leq I(A;V|U) \leq I(A;C|U) \leq I(A;C);  \nonumber \\
% \Delta & \leq R_C - H(C|A);  \nonumber \\
% R_A + \Delta &\geq H(A).   \nonumber
%\end{align}
%Hence, we eliminated the auxiliary random variables $U$ and $V$ and get the Slepian-Wolf region for $(R_A, R_C)$. It is also possible to see that the above bounds can be achieved with the usual binning scheme of Slepian-Wolf compression, therefore the above region is tight.

\section{Uncoded Side Information at Bob}\label{s:uncoded_SI}

A special case of Theorem \ref{t:SW_coded} is obtained when we have $R_C \geq H(C)$, that is, $C^N$ can be recovered by Bob with an arbitrarily small probability of error. Equivalently, we can assume that the side information sequence $B^N=C^N$ is available directly to Bob. The compression-equivocation rate region for this special case is given as below, which can be obtained from Theorem \ref{t:SW_coded}.

\begin{cor}\label{c:uncodedSI}
For uncoded side information $B^N$ at Bob, $(R_A, \Delta)$ is an achievable compression-equivocation rate pair if and only if,
\begin{eqnarray}
  R_A &\geq& H(A|B), \label{uncoded_A} \\
  0 \leq \Delta & \leq & [I(A;B|U) - I(A;E|U)]^+,  {~\rm and}  \label{uncoded_D} \\
  R_A + \Delta &\geq & H(A|E), \label{uncoded_T}
\end{eqnarray}
for some $U$ such that $U-A-(B,E)$ form a Markov chain.
\end{cor}

While Corollary \ref{c:uncodedSI} requires an auxiliary codebook generated by $U$ in the general case to conceal the source from the eavesdropper, in \cite{ITW08} we show that, when Bob's side information $B$ is \emph{less noisy} \cite{Csiszar_Korner} than Eve's side information $E$, Slepian-Wolf binning achieves the highest possible equivocation rate, i.e., (\ref{uncoded_D}) is maximized by a constant $U$. Furthermore, when Bob's side information is a stochastically degraded version of Eve's side information, no positive equivocation rate is achievable, and $\Delta=0$.

%
%Next, we consider a variation in the model of Corollary \ref{c:uncodedSI} in which Alice wants to transmit its information reliably to Bob who has access to $B^N$, while Eve is interested in some other correlated information denoted by $D^N$ rather than $A^N$. There is no requirement on the reconstruction of $D^N$ by Bob. This models a scenario in which a sensor transmitting its observation to the access point wants to limit the information leakage about some other correlated phenomenon that is security sensitive.
%
%\begin{cor}\label{c:mult_rx_1}
%$(R_A,\Delta)$ is achievable for the above scenario if and only if,
%\begin{eqnarray}
%  R_A &\geq& H(A|B), \nonumber \\
%  \Delta & \leq & \max \{H(D|A,B) + H(D|E,U) - H(D|B,U)\}, \nonumber \\
%  R_A + \Delta &\geq& H(A|E), \nonumber
%\end{eqnarray}
%in which we maximize over auxiliary random variables satisfying the Markov chain condition $U-A-(B,C,E)$.
%\end{cor}

In \cite{ITW08}, we also show that the availability of either Bob's or Eve's side information at Alice potentially increases the equivocation rate of the eavesdropper, while the compression rate bound on $R_A$ remains intact. When Alice does not have access to $B^N$, $\Delta =0$ if $B$ is independent of $A$. However, when $B^N$ is available to Alice, it is useful even if they are independent. This scenario is equivalent to Shannon's secret key model, in which a secure key, independent of the message, at rate $H(B)$ is shared by Alice and Bob.

\begin{lem}\label{lem:sec_key}
If $B$ is independent of $(A,E)$ and available to Alice as well, then $(R_A, \Delta)$ is achievable if and only if
\begin{eqnarray}
  R_A &\geq& H(A)  {~~\rm and} \nonumber \\
  0 \leq \Delta &\leq& \min \{H(B), H(A|E) \}. \nonumber
\end{eqnarray}
\end{lem}
\begin{proof}
The achievability follows by first compressing the source $A^N$ and then encrypting the compressed source bits with the secure shared key $B^N$, i.e., one time pad. The converse can be obtained from Corollary \ref{c:uncodedSI}.
\end{proof}

Now we compare two scenarios of having Eve's side information at Alice or Bob. When $E^N$ is available to Bob, the side information of Eve becomes physically degraded version of Bob's side information, and Slepian-Wolf compression would suffice \cite{ITW08}. The compression-equivocation rate region is characterized by
\begin{eqnarray}
  R_A &\geq& H(A|B,E) \nonumber \\
 0 \leq \Delta &\leq& I(A;B|E) {~~\rm and} \nonumber \\
  R_A  + \Delta &\geq& H(A|E). \nonumber
\end{eqnarray}
Note that, having $E^N$ at Bob helps decrease the compression rate bound on $R_A$ as well. Furthermore, as shown in \cite{ITW08} providing $E^N$ to Alice in addition to Bob would not help.

On the other hand, when $E^N$ is available to Alice, the compression-equivocation rate region is given by \cite{ITW08} $R_A \geq H(A|B)$, $0 \leq \Delta \leq I(A;B|E)$ and $R_A+\Delta \geq H(A|E)$. Comparing the two regions when $E^N$ is available to Alice or to Bob, we see that the latter requires a smaller compression rate due to a better side information at the receiver, while the equivocation rates are equal.

\section{Multiple Legitimate Receivers/Eavesdroppers}

Consider $K$ legitimate receivers, each with its own correlated side information $B_k$ for $k=1,\ldots,K$, that want to receive Alice's information reliably, while there is only one eavesdropper Eve. In the absence of an eavesdropper, a rate of $\max_{k} H(A|B_k)$ is necessary and sufficient for simultaneous reliable transmission to all the receivers \cite{Sgarro}.

From Corollary \ref{c:uncodedSI}, considering each receiver separately, the equivocation rate is bounded as $\Delta \leq \max \{H(A|E,U_k) - H(A|B_k,U_k)\}$ where the maximization is over $U_k$ satisfying the Markov chain $U_k - A- (B_k, E)$ for $k=1,\ldots,K$. The minimum of these individual equivocation rate bounds serves as an upper bound; however, achievability of it together with $R_A \geq \max_{k} H(A|B_k)$ does not follow directly. The achievability proof outlined in Appendix \ref{a:SW_coded} requires an auxiliary codeword to be decoded by the receiver. However, for multiple receivers, the auxiliary codebook $U_k$ that maximizes the equivocation rate for one of the users, might not be decodable by another user. Imposing such a decoding constraint requires a total transmission rate of $\max_{k} I(A;U_k|B_k) + \max_{k} H(A|B_k, U_k)$, which might be greater than $\max_{k} H(A|B_k)$, the required rate without the secrecy constraint.

Below, we give the compression-equivocation rate region in case of multiple receivers for two special cases. Proofs are omitted due to space limitation.
\begin{cor}\label{c:mult_rx_1}
If $A-B_k-E$ form a Markov chain for all $k=1,\ldots,K$, then $(R_A,\Delta)$ is achievable if and only if,
\begin{eqnarray}
  R_A &\geq& \max_{k} H(A|B_k), \nonumber \\
  \Delta & \leq & \min_{k} \{ H(A|E) - H(A|B_k) \} ,  {~\rm and} \nonumber \\
  R_A + \Delta &\geq& H(A|E). \nonumber
\end{eqnarray}
\end{cor}

\begin{cor}\label{c:mult_rx_2}
If $A-B_1-\cdots - B_K$ form a Markov chain, then $(R_A,\Delta)$ is achievable if and only if,
\begin{eqnarray}
  R_A &\geq& H(A|B_K), \nonumber \\
  \Delta & \leq & \max \{I(A, B_K| U) - I(A; E |U) \},  {~\rm and} \nonumber \\
  R_A + \Delta &\geq& H(A|E), \nonumber
\end{eqnarray}
where the maximization is over auxiliary random variables $U$ such that $U-A-(B_1,\ldots, B_K,E)$ form a Markov chain.
\end{cor}

In Corollary \ref{c:mult_rx_1}, due to the degradedness of $E$ with respect to $B_k$'s, picking a constant $U$ is optimal for all receivers. In Corollary \ref{c:mult_rx_2}, we use the auxiliary codebook $U$ that is chosen with respect to the worst receiver side information $B_K$.

Similarly, there may be multiple non-cooperating eavesdroppers all of which have their own correlated side information. Suppose there are $K$ eavesdroppers, the $k$-th of which has side information $E_k$. We have $K$ equivocation rates defined as \[\Delta_k \triangleq \frac{H(A^N|f_A(A^N), E_k^N)}{N} \mbox{ for } k=1,\ldots,K.\] This time, we pick the auxiliary codebook that simultaneously achieves the corresponding equivocation rates $\Delta_k$.

\begin{cor}\label{c:mult_rx_3}
$(R,\Delta)$ is achievable for the multiple eavesdropper scenario if and only if,
\begin{eqnarray}
  R_A &\geq& H(A|B), \nonumber \\
  \Delta_k & \leq & [H(A|E_k,U) - H(A|B,U)]^+,  {~\rm and} \nonumber \\
  R_A + \Delta_k &\geq& H(A|E_k), \nonumber
\end{eqnarray}
for $k=1,\ldots,K$ for some auxiliary random variable satisfying the Markov chain condition $U-A-(B,E_1,\ldots,E_K)$.
\end{cor}

\section{Conclusion}

In this paper, we have considered secure distributed compression in the presence of an eavesdropper. We have studied the case in which one of the transmitters is wire-tapped by an eavesdropper and we have shown that secure communication can be achieved with the help of the second transmitter who has its own correlated side information and a secure link to the legitimate receiver. We have provided inner and outer bounds to the compression-equivocation rate region for the model studied. We have also considered availability of side information at the transmitters, multiple legitimate receivers or multiple eavesdroppers. Future directions include extension to the lossy compression scenario.

\appendices
\section{Proof of Theorem \ref{t:SW_coded}}\label{a:SW_coded}

\emph{Inner bound:} We fix $p(u|a)$ and $p(v|c)$ satisfying the conditions in the theorem. Then we generate $2^{N(I(A;U)+\epsilon_1)}$ independent codewords of length $N$, $U^N(w_1)$, $w_1 \in \{1,\ldots,2^{N(I(A;U)+\epsilon_1)} \}$, with distribution $\prod_{i=1}^N p(u_i)$.  We randomly bin all $U^N(w_1)$ sequences into $2^{N(I(A;U|V)+\epsilon_2)}$ bins, calling them the auxiliary bins. For each codeword $U^N(w_1)$, we denote the corresponding auxiliary bin index as $a(w_1)$. On the other hand, we randomly bin all $A^N$ sequences into $2^{N(H(A|V,U)+\epsilon_3)}$ bins, calling them the source bins, and denote the corresponding bin index as $s(A^N)$. We also generate $2^{N(I(C;V)+\epsilon_4)}$  independent codewords $V^N(w_2)$ of length $N$, $w_2 \in \{1,\ldots,2^{N(I(C;V)+\epsilon_4)} \}$, with distribution $\prod_{i=1}^N p(v_i)$.

For each typical outcome of $A^N$, Alice finds a jointly typical $U^N(w_1)$. Then she reveals $a(w_1)$, the auxiliary bin index of $U^N(w_1)$, and $s(A^N)$, the source bin index of $A^N$, to both Bob and Eve; that is, the encoding function $f_A$ of Alice is composed of the pair $(a(w_1), s(A^N))$. Using standard techniques, it is possible to show that we have such a unique index pair with high probability. Charlie observes the outcome of its source $C^N$, finds a jointly typical $V^N(w_2)$ with $C^N$, and sends the index $w_2$ of $V^N$ over the private channel to Bob. With high probability there will be a unique $w_2$ such that $C^N$ and $V^N(w_2)$ are jointly typical.

Bob, having access to $w_2$ and the auxiliary bin index $a(w_1)$, can find the jointly typical $U^N(w_1)$ correctly with high probability. Then using $U^N$, the source bin index $s(A^N)$ and $V^N(w_2)$, Bob can reliably decode $A^N$. Since $H(C|A,V)=0$, knowing $A^N$ and $V^N$ correctly, Bob can find the correct $C^N$ with high probability as well. Letting $\epsilon_i \rightarrow 0$ for $i=1,2,3$ and $4$, we can make the total communication rate of Alice arbitrarily close to $I(A;U|V) + H(A|U,V) = H(A|V)$ and the rate of Charlie to $I(C;V)$. Since (\ref{SWrateC})-(\ref{SWrateA}) hold, these rates can be communicated to Bob while having arbitrarily small error probability for sufficiently large $N$.

The equivocation rate can be lower bounded as follow:
\begin{align}
& H(A^N|a(w_1), s(A^N), E^N) = H(A^N) - I(A^N; a(w_1), E^N) \nonumber \\
 &~~~~~~~~~~~~~~~~~~~~ - I(A^N; s(A^N) | E^N, a(w_1))  \nonumber\\
 & ~~~~~~~ \geq H(A^N) -  I(A^N; U^N, E^N) - H(s(A^N))  \label{ach1}
  \end{align}
 \begin{align}
  & ~~~~~~~ \geq H(A^N| U^N, E^N) - N H(A| V, U) - N\epsilon_3   \label{ach2}  \\
 & ~~~~~~~ = N [H(A| U, E) - H(A| V, U) -\epsilon_3]  \nonumber \\
 & ~~~~~~~ = N [I(A; V |U) - I(A; E |U) -\epsilon_3],  \nonumber
\end{align}
where (\ref{ach1}) follows from the data processing inequality; and (\ref{ach2}) follows from the fact that $s(A^N)$ is a random variable over a set of size $ 2^{N (H(A| V, U) + \epsilon_3)}$.

%The upper bound on equivocation rate is obtained as follows.
%\begin{align}
% & \frac{1}{N}H(A^N|a(w_1), s(A^N), E^N) \leq \frac{1}{N}H(A^N|a(w_1), s(A^N)) \nonumber \\
% & \leq \frac{1}{N} \left[ H(A^N|C^N, a(w_1), s(A^N)) \right. \nonumber \\
% & ~~~ \left.  + H(C^N| a(w_1), s(A^N)) - H(C^N| A^N, a(w_1), s(A^N)) \right]  \nonumber \\
% &\leq \frac{1}{N} H(C^N| a(w_1), s(A^N)) + \epsilon -H(C|A) \label{e:ach21} \\
% &= \frac{1}{N} \left[H(w_2| a(w_1), s(A^N)) + H(C^N| w_2, a(w_1), s(A^N)) \right. \nonumber \\
% & ~~~~~~~~ \left. - H(C|A) + \epsilon \right] \nonumber \\
% &\leq \frac{1}{N} H(w_2) + 2\epsilon - H(C|A) \label{e:ach22}  \\
% &\leq R_C  - H(C|A) + 2\epsilon \nonumber
%\end{align}
%where (\ref{e:ach21}) follows since $(a(w_1),s(A^N))-A^N-C^N$ form a Markov chain, and given $(C^N, a(w_1),s(A^N))$, $A^N$ can be determined with arbitrarily small probability of error; and (\ref{e:ach22}) follows similarly since given $(w_2,a(w_1),s(A^N))$, $C^N$ can be determined with arbitrarily small probability of error.

For $(U,V) \in \mathcal{P}_{in}$, we can show that
\begin{align}
I(A;V|U)-I(A;E|U) &\leq I(A;C) {~\rm and} \nonumber \\
I(A;V|U)-I(A;E|U) &\leq R_C - H(C|A)   \nonumber
\end{align}
Hence (\ref{SWrateZ}) is not active in the inner bound.

Finally, we also have
\begin{align}
& \frac{1}{N}H(A^N|a(w_1), s(A^N), E^N) \nonumber \\
 &= \frac{1}{N} \left[ H(A^N|E^N) - I(A^N; a(w_1), s(A^N)| E^N) \right]  \nonumber \\
&\geq H(A|E) - \frac{1}{N} H(a(w_1), s(A^N))  \nonumber \\
 &\geq H(A|E) - R_A.   \nonumber
\end{align}

\emph{Outer bound:} We define
\[J \triangleq f_A(A^N) \mbox{  and  } K \triangleq f_C(C^N).\] From Fano's inequality, we have
\begin{align}
H(A^N, C^N | J, K)\leq N \delta (P_e^N), \label{e:Fano}
\end{align}
where $\delta (\cdot)$ is a non-negative function with $\lim_{x \rightarrow 0} \delta(x) =0$.

Define $U_i \triangleq (J, A^{i-1},E^{i-1})$ and $V_i \triangleq (K, C^{i-1})$.  Note that both $U_i-A_i-(C_i,E_i)$ and $V_i-C_i-(A_i,E_i)$ form Markov chains. Then, we have the following chain of inequalities:
\begin{align}
N R_C \geq & H(K)  \nonumber \\
\geq & I(C^N; K)  \nonumber \\
= & \sum_{i=1}^N I(C_i;K|C^{i-1}) \label{ee:c11} \\
= & \sum_{i=1}^N I(C_i;K,C^{i-1}) = \sum_{i=1}^N I(C_i;V_i), \label{ee:c13}
\end{align}
where (\ref{ee:c11}) follows from the chain rule. We also have

\begin{align}
N R_A \geq& H(J) \geq H(J|K) \nonumber \\
 = & H(A^N, J| K) - H(A^N|J,K) \nonumber \\
 \geq & H(A^N|K) - N\delta(P_e^N) \label{ee:c21}\\
% = & \sum_{i=1}^N H(A_i|K,A^{i-1}) - N \delta(P_e^N) \nonumber \\
% \end{align}
% \begin{align}
 \geq & \sum_{i=1}^N H(A_i|K,A^{i-1}, C^{i-1}) - N \delta(P_e^N)  \label{ee:c22} \\
  = & \sum_{i=1}^N H(A_i|K,C^{i-1}) - N\epsilon  \label{ee:c23} \\
 = & \sum_{i=1}^N H(A_i|V_i) - N \delta(P_e^N),  \label{ee:c24}
\end{align}
where (\ref{ee:c21}) follows from (\ref{e:Fano}) and the nonnegativity of entropy; (\ref{ee:c22}) follows as $A_i-(K,A^{i-1})-C^{i-1}$ form a Markov chain; and (\ref{ee:c23}) follows as $A_i-(K,C^{i-1})-A^{i-1}$ form a Markov chain.

Next, we have the following set of inequalities:
\begin{align}
N \delta(P_e^N) \geq & H(C^N| J,K) \label{ee:c305} \\
\geq & H(C^N |A^N,K)  \label{ee:c31} \\
% = & \sum_{i=1}^N H(C_i|K, A^N, C^{i-1}) \label{ee:c32} \\
 = & \sum_{i=1}^N H(C_i|A_i, K, C^{i-1}) \label{ee:c33} \\
 = & \sum_{i=1}^N H(C_i|A_i, V_i),  \label{ee:c34}
\end{align}
where (\ref{ee:c305}) follows from (\ref{e:Fano}); (\ref{ee:c31}) follows since conditioning reduces entropy and $J$ is a function of $A^N$; and (\ref{ee:c34}) follows form the definition of $V_i$.

For the equivocation rate converse, we have
\begin{align}
N \Delta & =  H(A^N| J, E^N)  \nonumber \\
& =  H(A^N|J) - I(A^N;E^N|J) \nonumber \\
& =  H(A^N| J,K) + I(A^N; K|J) - I(A^N;E^N|J)  \nonumber \\
& \leq N\delta(P_e^N) + \sum_{i=1}^N I(A_i;K|J,A^{i-1}) - \sum_{i=1}^N H(E_i|J,E^{i-1}) \nonumber \\
& ~~~~~~ + H(E^N | A^N,J) \label{eq_1_3} \\
&\leq  N\delta(P_e^N) + \sum_{i=1}^N I(A_i;K|J,A^{i-1}, E^{i-1})  \nonumber \\
&~~~~~ -   \sum_{i=1}^N H(E_i|J,E^{i-1}, A^{i-1}) + H(E^N | A^N) \label{eq_1_4} \\
&\leq  N \delta(P_e^N) + \sum_{i=1}^N  \left[ I(A_i;K,C^{i-1}|J,A^{i-1}, E^{i-1})  \right. \nonumber \\
&~~~~~ - \left. H(E_i|J,E^{i-1}, A^{i-1}) + H(E_i | A_i) \right] \label{eq_1_4b} \\
&=  \sum_{i=1}^N \left[ I(A_i;V_i|U_i) - H(E_i|U_i) + H(E_i|A_i)  \right]+ N \delta(P_e^N) \label{eq_1_5} \\
&=   \sum_{i=1}^N \left[ I(A_i;V_i|U_i) - I(A_i; E_i|U_i) \right]+ N \delta(P_e^N) \label{eq_1_6}
\end{align}
where (\ref{eq_1_3}) follows from Fano's inequality and the chain rule; (\ref{eq_1_4}) follows from the memoryless property of the source and the side information sequences, and the fact that conditioning reduces entropy; (\ref{eq_1_4b}) follows from the chain rule and the non-negativity of the mutual information; (\ref{eq_1_5}) follows from definitions of $V_i$ and $U_i$; and finally (\ref{eq_1_6}) follows since $U_i-A_i-E_i$ form a Markov chain.

We also have
\begin{align}
N \delta(P_e^N) &\geq H(A^N, C^N |J, K) \label{eq:x1} \\
& = H(J,K|A^N, C^N) + H(A^N, C^N) -H(J,K) \nonumber\\
& \geq  H(A^N) + H(C^N) - I(A^N; C^N) -H(J) -H(K) \nonumber\\
& = H(A^N|J) - H(J|A^N) + NH(C|A) - H(K)  \nonumber \\
& \geq H(A^N| J, E^N) + NH(C|A) - N R_C \nonumber
\end{align}
where (\ref{eq:x1}) follows from (\ref{e:Fano}). We get
\begin{align}
\delta(P_e^N) &\geq \Delta + H(C|A) - R_C, \label{eq:xx}
\end{align}
and
\begin{align}
N \delta(P_e^N) &\geq H(A^N, C^N |J, K)  \nonumber \\
& = H(A^N, C^N) - I(A^N, C^N; J,K) \nonumber \\
& = H(A^N, C^N) - H(A^N) + H(A^N|J) - H(C^N) \nonumber \\
& ~~~~ + H(C^N|K) + I(K;J)  \nonumber\\
& \geq H(A^N|J, E^N) - I(A^N; C^N) \nonumber \\
& \geq N\Delta  - NI(A;C). \nonumber
\end{align}

And finally we have
\begin{align}
H(A|E) &\leq \frac{1}{N} H(A^N, J|E^N)  \nonumber\\
& = \frac{1}{N} \big[ H(J|E^N) + H(A^N|E^N, J)  \big] \nonumber\\
& \leq \frac{H(J)}{N} + \Delta \nonumber\\
& \leq R_A + \Delta. \label{eq:xxx}
\end{align}

Now, we define a new independent random variable $Q$ uniformly distributed over the set $\{1,2,\ldots,N\}$, and $A \triangleq A_Q$, $E \triangleq E_Q$, $V \triangleq (V_Q, Q)$, and $U \triangleq (U_Q,Q)$. Letting $N \rightarrow \infty$ and $P_e^N \rightarrow 0$ we obtain the outer bound in the theorem.

\end{document}